\documentclass{pasj00}

\SetRunningHead{K.~Saitou \etal}{Suzaku Observation of Z~Cam}

\Received{2011/12/02}
\Accepted{2012/02/27}

\begin{document}

\title{Suzaku X-Ray Observation of the Dwarf Nova Z~Camelopardalis\\
       at the Onset of an Optical Outburst}

\author{%
Kei~\textsc{Saitou},\altaffilmark{1,2}
Masahiro~\textsc{Tsujimoto},\altaffilmark{1}
Ken~\textsc{Ebisawa},\altaffilmark{1,2}
and
Manabu~\textsc{Ishida}\altaffilmark{1}
}
\altaffiltext{1}{%
Japan Aerospace Exploration Agency, Institute of Space and Astronautical Science,\\
3-1-1 Yoshinodai, Chuo-ku, Sagamihara, Kanagawa 252-5210}
\altaffiltext{2}{%
Department of Astronomy, Graduate School of Science, The University of Tokyo,
7-3-1 Hongo, Bunkyo-ku, Tokyo 113-0033}
\email{ksaitou@astro.isas.jaxa.jp}


\maketitle

\begin{abstract}
We present the result of a Suzaku X-ray spectroscopic observation of the dwarf nova
Z~Camelopardalis, which was conducted by chance at the onset of an optical outburst.
We used the X-ray Imaging Spectrometer (a 38~ks exposure) and the Hard X-ray Detector
(34~ks) to obtain a 0.35--40~keV spectrum simultaneously.
Spectral characteristics suggest that the source was in the X-ray quiescent state
despite being in the rising phase of an outburst in the optical band.
The spectrum shows a clear signature of circumstellar absorption in excess of
interstellar absorption and the reprocessed emission features of Fe fluorescence and
Compton scattering.
The extra absorption is explained due to partial coverage by either neutral or ionized
matter.
We found a spectral change during the observation, which is attributable only to the
change in the circumstellar absorption. 
Such an X-ray spectral variation is reported for the first time in dwarf novae.
We speculate that the variation in the circumstellar absorption is interpreted as
a time-varying disk wind or geometrically flaring disk around the white dwarf during
the propagation of a heat wave inward along the accretion disk at the beginning of
the outburst, in which optical outburst and X-ray quiescent states co-exist.
\end{abstract}

\section{Introduction} \label{s1}
Dwarf novae (DNe) are a subclass of non-magnetic cataclysmic variables, which are
close binary systems consisting of a white dwarf (WD) and a late-type companion star
\citep{Warner1995,Hellier2001}.
The defining characteristic of DNe is repeated optical outbursts, in which the optical
brightness increases by 2--5~mag.
Thermal-viscous instability in the accretion disk is considered to be the cause
(e.g., \cite{Smak1984,Osaki1996,Lasota2001}). 
The cycle repeats between two distinctive stable disk states: the quiescent state
with the disk mostly made of neutral hydrogen and the outburst state mostly made of
ionized hydrogen.
The viscosity, surface density, and the mass accretion rate are higher in the latter
state.

DNe are known to be X-ray emitters.
The X-ray features typically show a dichotomy between the quiescent and outburst states.
In the quiescent state, on one hand, the X-rays are from the hot and optically-thin
thermal plasma, which is produced in the boundary layer between the inner edge of the
accretion disk rotating at a Keplerian velocity and the WD surface rotating at a slower
spin period.
The plasma is localized within a height of $\sim$$0.1R_{\mathrm{WD}}$ from the surface
of the WD with a radius of $R_{\mathrm{WD}}$ \citep{Mukai1997}.
The X-ray spectrum is very hard with a temperature beyond 10~keV and strong Fe~K
emission features.
In the outburst state, on the other hand, the hard emission is suppressed by a factor
of a few and is replaced by extreme ultra-violet (EUV) plasma emission with a much
lower temperature.
This is because the plasma in the boundary layer becomes optically thicker and more
efficient in radiative cooling due to the increased accretion rate to the WD.

One of the features often seen in the X-ray spectra of DNe is a complex extinction
structure in the soft band ($<$2~keV).
The extinction cannot be explained by interstellar absorption alone, which indicates
the presence of circumstellar absorption.
In eclipsing systems with high inclination angles (e.g., OY~Car: \cite{Ramsay2001} and
V893~Sco: \cite{Mukai2009}), the complex extinction can be naturally interpreted as a
partial absorption by a part of the accretion disk.
However, the extra extinction is also seen in some non-eclipsing systems with lower
inclination angles (e.g., SS~Cyg: \cite{Done1997} and Z~Cam: \cite{Baskill2001}).
One idea to explain such extinction is disk wind, which is ionized to some extent
and intervenes partially or fully the line of sight. 

Disk winds are a phenomenon seen in any types of compact objects with an accretion
disk (e.g., \cite{Ueda2001,Boirin2003,Kubota2007,Tombesi2010}).
They may play an important role in the feedback process of energy and matter from
compact objects to the interstellar space and also to the intergalactic space in case
of active galactic nuclei (e.g., \cite{Elvis2006,Fabian2010}).
In nearby DNe, which are accessible with optical and ultra-violet (UV) spectroscopic
observations, disk wind is observed in many systems as P~Cygni profiles
\citep{Robinson1973,Cordova1982,Klare1982,Szkody1986}, motivating the progress in
spectral synthesis modeling of disk winds
\citep{Shlosman1993,Knigge1995,Feldmeier1999,Long2002}.
If we can trace features of disk wind in X-ray spectra of DNe, which is yet to be
established, it will bring a wider application to constrain the wind parameters
in a larger number of objects in a variety of phases in the quiescent and outburst
cycles.
Of particular interest is the transition phase between the two states, in which we may
be able to distinguish the extinction by disk wind from other causes, as we expect the
disk wind to change its strength and structure during the transition.

\medskip

In this paper, we present the result of a Suzaku X-ray observation of Z~Cam, which
was conducted by chance at the onset of an optical outburst. 
As we will discuss in \S~\ref{s5-2}, the source was in the X-ray quiescent phase, despite
being in the outburst phase in the optical, suggesting that the heat wave had not
reached the boundary layer. 
We utilize this opportunity to investigate the presence of circumstellar absorption in
the X-ray spectrum and examine any changes in the feature.
Only a few X-ray observations were made to date in this particular phase of DN outbursts
(e.g., Z~Cam: \cite{Baskill2001}; SS~Cyg: \cite{Wheatley2003}). 

The plan of this paper is as follows.
In \S~\ref{s2}, we briefly summarize the basic properties of Z~Cam with an emphasis on
disk winds.
In \S~\ref{s3} and \S~\ref{s4}, we describe our observation and the result of temporal
and spectral analysis, in which we present the presence of the circumstellar absorption
and its time variation.
We discuss some possible interpretations of our finding in \S~\ref{s5}, and conclude
in \S~\ref{s6}.

\section{Object --- Z~Camelopardalis} \label{s2}
Z~Cam is the archetype of the Z~Cam subgroup of DNe.
It is one of the brightest DNe in the optical band ($V$=10.5--13.0~mag) and thus is
a well studied object.
Major parameters are summarized in table~\ref{t1}.

Sources in the Z~Cam subgroup show an optical outburst about every few months
($\sim$26~days on average for Z~Cam; \cite{Oppenheimer1998}). 
The defining characteristic of this subgroup is the ``stand-still'' phase after some
outbursts.
For a certain period of time, the brightness stays in the middle of the outburst peak
and the quiescence level, rather than decaying monotonically to the quiescence level.
The stand-still phase continues for several days to even years
(e.g., \cite{Oppenheimer1998}).

There is ample evidence that Z~Cam has a mass loss in the form of wind.
The object is thus used as a test bed for constructing UV spectral synthesis models of
disk wind.
The spatial structure of the wind is investigated by various authors
\citep{Knigge1997,Long2002,Hartley2005}, in which rotating biconical wind was found to
explain various observed features very well.
It is known that the disk wind appears and disappears depending on the phase of the
outburst and quiescent cycle.
The P~Cygni profile of the C\emissiontype{IV} feature, which is the most prominent
feature of the disk wind, was seen in the decline phase from an outburst, and one month
from the start of a stand-still phase, but not in six months from the start of a
stand-still and at quiescence \citep{Szkody1986,Knigge1997,Hartley2005}.
With a poor sampling only with UV observations, a picture of the temporal behavior of
the wind; i.e., when the wind starts and stops emanating from the disk, is not yet
clear. 

Z~Cam was observed in the X-ray band several times with the EXOSAT \citep{Mukai1993},
ROSAT \citep{Wheatley1996}, and ASCA \citep{Baskill2001} observatories.
A hint of X-ray absorption by wind was seen in two ASCA observations \citep{Baskill2001},
in which extra absorption upon the interstellar absorption was required to explain the
X-ray spectra.
The amount of the extra absorption, which was successfully modelled by ionized absorber,
was larger than the interstellar absorption by two orders of magnitude in both
observations taken during an optical outburst and a transition phase from quiescence to
the outburst.
However, due to limited statistics and resolution in the soft energy band, a possible
time variation in the circumstellar absorption was not detected between the two
observations and within each observation.

\begin{table*}[t]
\begin{center}
  \caption{Parameters of the Z~Cam system.}
  \label{t1}
\begin{tabular}{lllll}
  \hline
  Parameter & Value & Unit & Method & Ref.\footnotemark[$*$] \\
  \hline
  Distance ($d$)
    & $163_{-38}^{+68}$ & pc & Trigonometric parallax
    & [1] \\
  Interstellar absorption ($N_{\mathrm{H}}^{\mathrm{ISM}}$)
    & $4\times 10^{19}$ & cm$^{-2}$ & UV spectroscopy
    & [2--5] \\
  Inclination angle ($i$)
    & $68$ & deg & Far-UV spectroscopy
    & [6] \\
  Orbital period ($P_{\mathrm{orb}}$)
    & $6.96$ & hour & Optical photometry and spectroscopy
    & [7] \\
  WD mass ($M_{\mathrm{WD}}$)
    & $0.99\pm 0.15$ & $\MO$ & Radial velocity
    & [8] \\
  WD radius ($R_{\mathrm{WD}}$)
    & $5.8_{-1.2}^{+1.1}\times 10^{8}$ & cm & Mass-radius relation
    & [6, 9] \\
  Companion mass ($M_{2}$)
    & $0.70\pm 0.15$ & $\MO$ & Radial velocity
    & [4, 8] \\
  Companion spectral type
    & K7 & & Optical spectroscopy
    & [10] \\
  \hline
  \multicolumn{5}{@{}l@{}}{\hbox to 0pt{\parbox{155mm}{\footnotesize
  \par \noindent
  \footnotemark[$*$]
    References are --- 
    [1]~\citet{Thorstensen2003}; [2]~\citet{Mauche1988}; [3]~\citet{Wheatley1996};
    [4]~\citet{Knigge1997}; [5]~\citet{Baskill2001}; [6]~\citet{Hartley2005};
    [7]~\citet{Thorstensen1995}; [8]~\citet{Shafter1983}; [9]~\citet{Anderson1988};
    [10]~\citet{Ritter2003}.
  }\hss}}
\end{tabular}
\end{center}
\end{table*}
%

\section{Observation and Data Reduction} \label{s3}
\begin{figure}[t]
\begin{center}
  \FigureFile(80mm,80mm){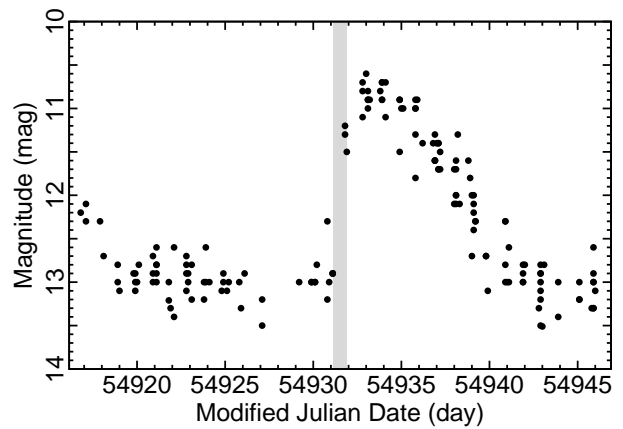}
\end{center}
  \caption{Optical light curve of Z~Cam taken from the AAVSO.
           The duration of the Suzaku observation is represented by the shaded
           region.}
  \label{f1}
\end{figure}

We observed Z~Cam with the Suzaku satellite \citep{Mitsuda2007} on 2009 April 10. 
Figure~\ref{f1} shows the optical light curve covering our X-ray observation taken
from the American Association of Variable Star Observers (AAVSO)\footnote{%
  See http://www.aavso.org/ for details.
}.
The X-ray observation was made during the onset of an optical outburst for a total
duration of 65~ks, corresponding to 2.6~orbits of Z~Cam.

Suzaku has two operating instruments covering different energy ranges.
One is the X-ray Imaging Spectrometer (XIS: \cite{Koyama2007}) sensitive at an
energy range of 0.2--12~keV. 
The XIS is equipped with four X-ray CCD cameras, three of which (XIS~0, 2, 3) are 
front-illuminated (FI) devices and the remaining one (XIS~1) is a back-illuminated
(BI) device.
The FI and BI devices show a superior response to each other in the hard and the
soft band, respectively.
The XIS~2 is dysfunctional since 2009 November, we thus used the remaining devices
in this paper.
In combination with four X-ray telescopes \citep{Serlemitsos2007} co-aligned with
each other, the XIS has an imaging capability to cover a $17\farcm 8\times 17\farcm 8$
field of view (FoV) with a pixel scale of $1\farcs 04$~pixel$^{-1}$ and a telescope
half-power diameter of $\sim$$2\farcm 0$.
A total effective area for the remaining three CCDs is 1030~cm$^{2}$ at 1.5~keV.
The energy resolution in the full width at a half maximum (FWHM) is 170--220~eV at
5.9~keV as of the observation date.

The other instrument is the Hard X-ray Detector (HXD:
\cite{Takahashi2007,Kokubun2007,Yamada2011}) sensitive at an energy range of
10--600~keV.
The HXD consists of PIN diodes and GSO scintillators, which together compose
a non-imaging detector.
We concentrate on the PIN data at 10--70~keV in this paper.
An effective area of the PIN is $\sim$160~cm$^{2}$ at 20~keV.
The passive fine collimators restrict the FoV to $\sim$34$\arcmin$ square in the
FWHM and $\sim$70$\arcmin$ square in the full width at zero intensity (FWZI).
The narrow FoV, the surrounding anti-coincidence detectors, and the low and stable
instrumental background enable us to achieve unprecedented sensitivity in this
energy band. 

In our observation, the target was placed at the center of the XIS FoV.
The XIS was operated in the normal clocking mode with a frame time of 8~s.

The data were screened by the standard pipeline processing version~2.3, in which
we discarded events during South Atlantic Anomaly passages and Earth elevation
angles below 5~degrees.
For the XIS, events were further removed during the elevation angles from the
sun-lit Earth below 20~degrees.
For the PIN, events taken at the cut-off-rigidity of less than 6~GV were removed.
As a result, we obtained a net exposure time of 38~ks for the XIS and 34~ks for
the PIN.
Throughout this paper, we used the HEASoft\footnote{%
  See http://heasarc.nasa.gov/docs/software/lheasoft/ for details.
} version~6.8 for data reduction and the \texttt{Xspec} package version~12.7
\citep{Arnaud1996} for spectral fitting. 

%

\section{Analysis} \label{s4}

\subsection{Event Extraction} \label{s4-1}
In the XIS image, we see no source besides Z~Cam.
We accumulated the source events from a circular region of $3\farcm 13$ (180~pixels)
in radius, which maximizes the signal-to-noise ratio. 
The background events were extracted from an annulus of 4$\arcmin$--7$\arcmin$ radii
concentric to the source region. 
The encircled energy fraction of the background region is approximately 3\% of the
source region. 

For the PIN, which is a non-imaging detector, the background consists of the
instrumental non-X-ray background (NXB), the cosmic X-ray background (CXB), and
possible contaminating sources within the FoV. 
We used NXB events provided by the instrument team \citep{Fukazawa2009} and simulated
CXB events assuming a model obtained with the HEAO-1 satellite \citep{Boldt1987}. 
We checked the latest INTEGRAL IBIS \citep{Bird2010} and Swift BAT \citep{Cusumano2010}
catalogues and found no contaminating source within the FWZI FoV of the PIN.

\subsection{Temporal Analysis} \label{s4-2}

%
\begin{figure*}[t]
\begin{center}
  \FigureFile(140mm,80mm){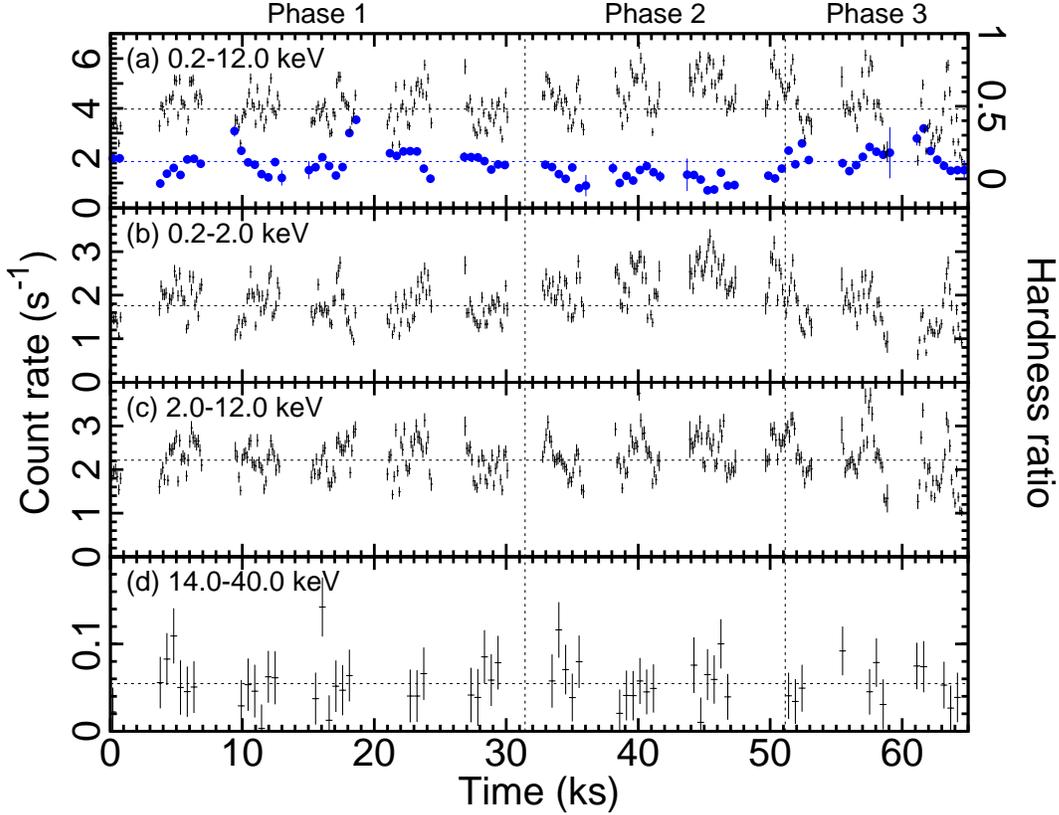}
\end{center}
  \caption{Light curves of background-subtracted count rates in the (a)~0.2--12~keV,
           (b)~0.2--2.0~keV, and (c)~2.0--12.0~keV bands using the XIS and
           (d)~14--40~keV band using the PIN.
           The light curve of the hardness ratio in the XIS band is overlaid in (a).
           The curves are binned with 128~s for the XIS count rate curves,
           512~s for the PIN count rate curve and the XIS hardness ratio curve.
           The discontinuities are due to Earth occultation of the object.
           A 1$\sigma$ Poisson statistical uncertainty is given.
           The three phases are divided by the dotted vertical lines.
           The dotted horizontal lines indicate the mean count rate and the mean
           hardness ratio derived from phase~1.
           The origin of the abscissa is the time of the observation start at 
           MJD$=54931.147$~d.}
  \label{f2}
\end{figure*}

Using the XIS (0.2--12~keV) and PIN (14--40~keV) data, we constructed light curves of
the background-subtracted count rate (figure~\ref{f2}).
We constructed the curves with several different time bins to find any changes in a
wide range of time scales.
As a result, we found two apparent changes of different time scales.

One is a fluctuation in the XIS count rate in a short time scale of $\sim$100~s, which
was found most prominently in the light curve binned with 16~s.
The count rate changes by $\sim$50\% in a range of 2--8~s$^{-1}$ with a mean of
4.1~s$^{-1}$ in this bin size.

The other is an increase in the XIS count rate at around the $\sim$30--50~ks interval
from the start of the observation (figure~\ref{f2}a).
In order to investigate this change further, we also constructed band-limited curves
in the soft (0.2--2~keV) and the hard (2--12~keV) band of the XIS (figure~\ref{f2}b and
c, respectively), as well as the hardness ratio (figure~\ref{f2}a; blue curve) defined
as $(H+S)/(H-S)$, where $H$ and $S$ are the count rate in the hard and the soft band,
respectively.
The change in the $\sim$30--50~ks interval is significantly seen only in the soft band,
which is confirmed both with the count rate and the hardness ratio curves.
No similar change is seen in the PIN light curve (figure~\ref{f2}d).
For time-sliced spectroscopy in later sections (\S~\ref{s4-4}), we divided the observation
time into three phases; phase~2 being the interval with a significant decrease in the
hardness ratio and the phases~1 and 3 being the intervals prior and posterior to the
phase~2, respectively (figure~\ref{f2}).

\begin{figure}[t]
\begin{center}
  \FigureFile(80mm,80mm){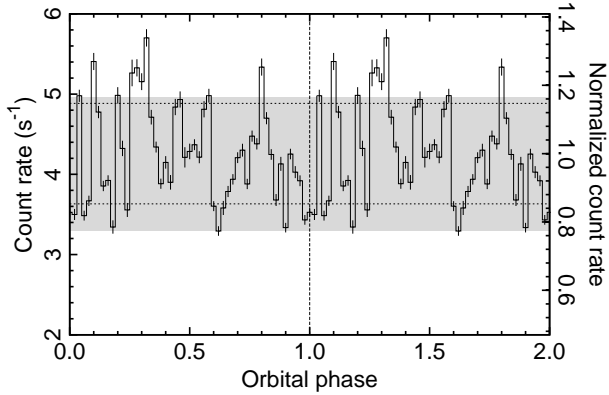}
\end{center}
  \caption{Light curve folded by the orbital period in 0.2--12~keV with a bin size of
           $\sim$500~s~bin$^{-1}$.
           A 1$\sigma$ Poisson statistical uncertainty is given for the data.
           Two orbital phases are shown for clarity.
           The orbital phase~0.0 corresponds to the observation start. 
           The vertical ticks on the left-hand side represent the count rate, while those
           on the right-hand side represent the count rate normalized by the mean count
           rate.
           The standard deviation of the count rate is shown with the horizontal dotted
           lines for the folded light curve and with the gray bands for the unfolded
           light curve.}
  \label{f3}
\end{figure}

We also investigated changes associated with the orbital period of $P_{\mathrm{orb}}=25.0$~ks.
Figure~\ref{f3} shows a count rate curve folded by the period, in which no apparent
variation was found except for the short time fluctuation seen also in the light curve
before folding (figure~\ref{f2}).
To quantify the claim, we compared the mean count rate and its standard deviation between
the folded and unfolded curves and found that they are consistent with each other.
In fact, similar fluctuation was seen in light curves folded by any other arbitrary
chosen periods.
Also, the aforementioned feature of the hardness change in the 30--50~ks interval was
seen only once and did not repeat itself.
We thus conclude that there is no X-ray change associated with the orbital period.

%

\subsection{Spectral Analysis (1): Time-Averaged Spectrum} \label{s4-3}
We first present the spectral fitting for the time-averaged spectrum. 
We constructed background-subtracted spectrum using the XIS and the PIN.
For the XIS, we generated the detector and telescope response files using the
\texttt{xisrmfgen} and \texttt{xissimarfgen} \citep{Ishisaki2007} tools, respectively.
The two FI spectra were merged for their nearly identical response, while the BI
spectrum was treated separately. 
The 1.8--2.0~keV band was removed for a known calibration uncertainty at the Si~K edge.
At the soft-band end of the response, which is affected by accumulating contamination
material on the surface of the CCDs, the calibration has shown some progress recently
by introducing time variability in the chemical composition of the contaminants.
We examined the spectra of two calibration sources, 1E~0102.2--7219 and RX~J1856.5--3754,
that are closest in time with ours and found that the systematic uncertainty in the
lowest end of the response is $\lesssim$20\% at 0.35~keV from the deviation of the
response-convolved model to the data.
For the PIN, we used the standard detector response file distributed by the instrument
team.

\begin{figure}[t]
\begin{center}
  \FigureFile(80mm,80mm){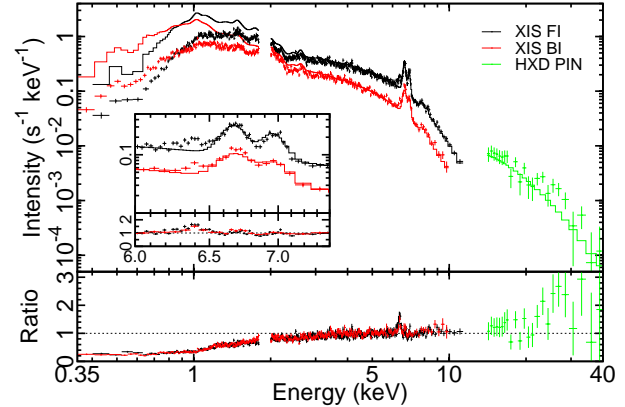}
\end{center}
  \caption{Background-subtracted and time-averaged spectrum fitted intentionally with
           a simple model of multi-temperature thin-thermal plasma emission
           attenuated by the absorption fixed to the interstellar value.
           The 4--10~keV data were used for the fitting.
           The data (crosses) and the best-fit model (solid lines) are shown in the
           upper panel, while the ratio of the data to the best-fit model is in the
           lower panel.
           An enlarged view at the Fe~K$\alpha$ complex band is shown in the inset.}
  \label{f4}
\end{figure}

Figure~\ref{f4} shows the 0.35--40~keV spectrum using both the XIS and the PIN.
The most prominent feature in the entire spectrum is the intense emission lines at
6.7 and 7.0~keV respectively from the $n=2\rightarrow 1$ lines by Fe\emissiontype{XXV} 
and the Ly$\alpha$ line by Fe\emissiontype{XXVI}, which indicates an optically-thin
thermal plasma. Other features can also be noticed.

In order to characterize the spectrum, we began with a simple model; i.e.,
multi-temperature thin-thermal plasma model (\texttt{cemekl}) attenuated by photoelectric
absorption (\texttt{tbabs}: \cite{Wilms2000}) of a column fixed to the interstellar value
(table~\ref{t1}).
The multi-temperature model is an integral of a single-temperature plasma model
(\texttt{mekal}: \cite{Mewe1985,Mewe1986,Kaastra1992,Liedahl1995}) with 
a differential emission measure ($EM$) as a function of the plasma temperature ($T$)
in a power-law form of 
$d(EM)/dT\propto (T/T_{\mathrm{max}})^{\alpha-1}$.
Such a model is widely used to describe X-ray spectra from CVs (e.g.,
\cite{Done1997}; \cite{Baskill2005}).
The free parameters were the maximum temperature ($T_{\mathrm{max}}$) and the power
($\alpha$) in the distribution as well as the abundance ($Z$) changed collectively for
all metals relative to hydrogen with respect to the solar value.
We employed the solar abundance by \citet{Wilms2000} and the photo-ionization
cross-sections by \citet{Verner1996}.
The 4--10~keV data were used for the fitting, and the best-fit model was extrapolated
both to the lower and upper energy ranges (figure~\ref{f4}). 
We multiplied a constant of 1.164 to the PIN normalization against the XIS to compensate 
for the known systematic difference between the two\footnote{%
  See a Suzaku memo
  (http://www.astro.isas.jaxa.jp/suzaku/doc/\\suzakumemo/suzakumemo-2008-06.pdf)
  for details.
}. 
We call this model a ``fiducial model''.

The residuals to the fiducial model (figure~\ref{f4}) indicate that the spectral
model requires other components to account for the following:
(1)~excess emission line at 6.4~keV presumably from Fe\emissiontype{I} K$\alpha$
fluorescent emission,
(2)~extra attenuation below $\sim$2~keV, which is attributable to extinction by
circumstellar medium (CSM), and
(3)~excess emission above $\sim$20~keV, which is a signature of Compton-scattered
continuum emission.
In the given spectrum, we found that these multiple spectral components are coupled
with each other.
Therefore, we started by inspecting each component separately by local fitting in
carefully selected energy bands (\S~\ref{s4-3-1}--\S~\ref{s4-3-3}) and then conducted
the entire fitting with a synthesized model (\S~\ref{s4-3-4}).

%

\subsubsection{Fe fluorescent line} \label{s4-3-1}
First, we constrained the emission line at 6.4~keV.
We fitted the 4--10~keV spectrum with the fiducial model plus a Gaussian line component.
The center energy and the intensity of the line were the free parameters, while the intrinsic width
was fixed to 0~eV in the fitting.
We obtained the best-fit center energy of $6.40_{-0.01}^{+0.02}$~keV, which is consistent
with the K$\alpha$ fluorescence line from Fe\emissiontype{I}.
We fixed the center energy to this value in the following steps.

\subsubsection{Reflection component} \label{s4-3-2}
\begin{figure}[t]
\begin{center}
  \FigureFile(80mm,80mm){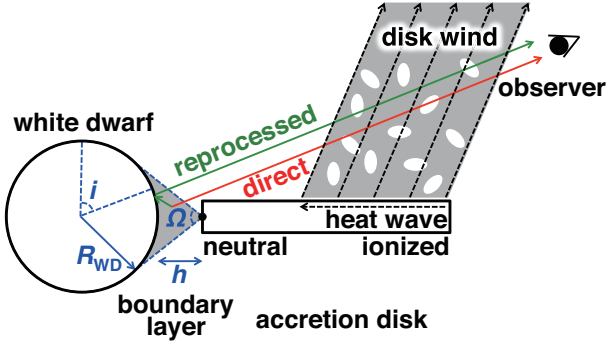}
\end{center}
  \caption{Schematic view of the plasma structure and the disk wind.
           The disk wind is illustrated for only the upper-right part of the
           accretion disk.}
  \label{f5}
\end{figure}

Second, we constrained the excess continuum emission by Compton scattering in the
PIN band.
The fluorescence and the Compton scattering are coupled physically under the same 
geometry with a common parameter $\Omega /2\pi$; the solid angle subtended by the
reflector viewed from the plasma (figure~\ref{f5}).
However, because no spectral model is available to account for both processes in
\texttt{Xspec}, we iterated the fitting procedure until we obtained a converged
result between the two.
The relation between the equivalent width (EW$_{\mathrm{Fe\emissiontype{I}}}$) of the
Fe fluorescence line and the viewing angle ($\Omega /2\pi$) was taken from
\citet{George1991}.
The inclination angle was fixed to the value in table~\ref{t1}.

We fitted the 4--40~keV spectrum using the fiducial model plus a Gaussian line
component, which is modified by the Compton reflection model (\texttt{reflect}:
\cite{Magdziarz1995}).
We started with $\Omega /2\pi =1$ as an input parameter for the Compton reflection
model.
The derived EW$_{\mathrm{Fe\emissiontype{I}}}$ was compared with the value expected to
constrain the viewing angle $\Omega /2\pi$.
In \citet{George1991}, a power-law spectrum is assumed for the photo-ionization
continuum emission.
We approximated the best-fit thermal model with a power-law model in the 7.11--10~keV
band, and derived the power-law slope.
We also corrected for the Fe abundance [Fe/H]; we used $2.69\times 10^{-5}$, while
\citet{George1991} assumed $3.3\times 10^{-5}$.
The derived $\Omega /2\pi$ value was used as an input parameter of the Compton
reflection model in the next iteration.
This process was repeated until we obtained a converged value of $\Omega /2\pi =0.484$.
The parameter is fixed to this value in the following steps. 

%

\subsubsection{Additional absorber} \label{s4-3-3}
\begin{figure}[t]
\begin{center}
  \FigureFile(80mm,80mm){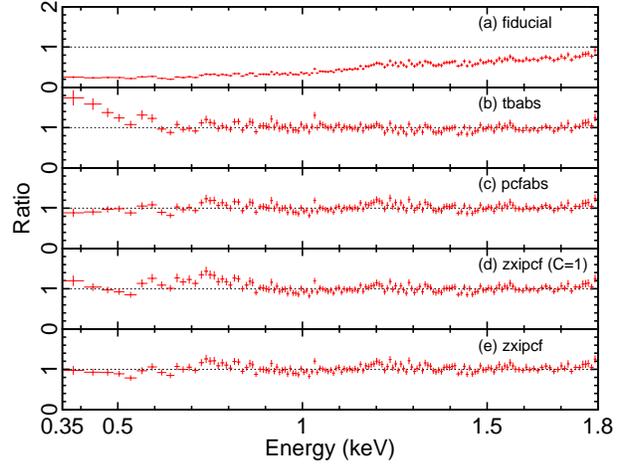}
\end{center}
  \caption{Ratio of the data against the models with different circumstellar absorption
           models:
           (a)~the fiducial model (no absorption by CSM),
           (b)~the \texttt{tbabs} model,
           (c)~the \texttt{pcfabs} model,
           (d)~the \texttt{zxipcf} model with the covering fraction $C^{\mathrm{CSM}}=1$,
           and 
           (e)~the \texttt{zxipcf} model. 
           The BI spectrum is displayed.}
  \label{f6}
\end{figure}

Third, we constrained the additional extinction prominent below $\sim$2~keV
(figure~\ref{f4}).
We convolved the model described above with several additional absorption models
and compared their reduced $\chi^{2}$ values ($\chi_{\mathrm{red}}^{2}$) in the
0.35--1.8~keV band (figure~\ref{f6}).

For the additional model, we started with the photoelectric absorption by neutral
matter using the same model with the interstellar absorption (\texttt{tbabs}). 
This model was unsuccessful with $\chi_{\mathrm{red}}^{2}=1.90$ for 309 degrees of
freedom (dof).

The residual (figure~\ref{f6}b) indicates that the observed spectrum remains less 
attenuated than the model predicts despite the fact that the deviation from the
fiducial model (figures~\ref{f4} and \ref{f6}a) starts at an energy as high as
$\sim$3~keV.
This is a signature of either or both of the partial coverage by the absorber or the
absorber being ionized.
We thus employed a partial absorption by neutral matter (\texttt{pcfabs}), a full
absorption by ionized matter (\texttt{zxipcf} with the covering fraction fixed to 1),
or a partial absorption by ionized matter (\texttt{zxipcf}, in which the covering fraction
is a free parameter).
The \texttt{zxipcf} model is calculated using the \texttt{XSTAR} code
\citep{Bautista2001}.
The free parameters in the \texttt{pcfabs} model are the absorption column
($N_{\mathrm{H}}^{\mathrm{CSM}}$) and the covering fraction ($C^{\mathrm{CSM}}$), while
those of the \texttt{zxipcf} are $N_{\mathrm{H}}^{\mathrm{CSM}}$, $C^{\mathrm{CSM}}$, and
the ionization parameter ($\xi^{\mathrm{CSM}}$).
There is a systematic deviation in the residual around 0.7--0.8~keV for
the full-covering ionized model (figure~\ref{f6}d), while no such deviation is seen in
the entire soft-band spectrum for the partial-covering neutral and ionized models
(figure~\ref{f6}c and e).
Indeed, the latter two show a better $\chi_{\mathrm{red}}^{2}$ values of $1.46$
(dof~$=308$) and $1.47$ (dof~$=307$), respectively, than the former of $1.79$
(dof~$=308$). 
We therefore used partially-covering neutral or ionized matter for the additional
absorption in the following steps. 

%

\subsubsection{Entire fitting} \label{s4-3-4}
\begin{figure}[t]
\begin{center}
  \FigureFile(80mm,80mm){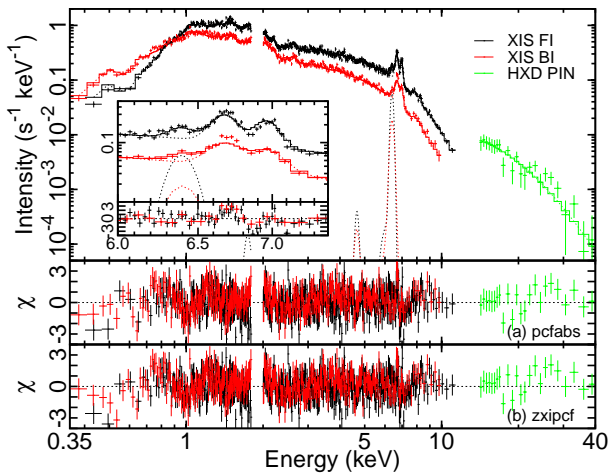}
\end{center}
  \caption{Background-subtracted time-averaged spectra with the best-fit models.
           The symbols follow figure~\ref{f4}.
           Each spectral components are represented by dotted lines.
           The two residuals are for the fitting with partial coverage by 
           (a)~neutral matter and 
           (b)~ionized matter for the additional absorption.
           The former model is used for the uppermost panel.}
  \label{f7}
\end{figure}

Finally, we combined all the aforementioned components to fit the entire spectrum
in the 0.35--40~keV band.
To the fiducial model, we added a Gaussian line and convolved with the Compton
reflection model and one of the two additional absorption models: partial absorption 
by neutral or ionized matter. 
We conducted the iteration process described in \S~\ref{s4-3-2} to reach a consistent
geometrical solution between the fluorescence and the Compton scattering.
We obtained a successful fit for the two models of the additional absorption, which
are shown in figure~\ref{f7} and tables~\ref{t2} and \ref{t3}.

\begin{table*}[t]
\begin{center}
  \caption{Best-fit parameters for the neutral partial covering absorption model.\footnotemark[$*$]}
  \label{t2}
\begin{tabular}{lccccccccc}
  \hline
  Phase
    & $N_{\mathrm{H}}^{\mathrm{ISM}}/10^{21}$
    & $N_{\mathrm{H}}^{\mathrm{CSM}}/10^{21}$ & $C^{\mathrm{CSM}}$
    & $\Omega /2\pi$ 
    & $\alpha$ & $kT_{\mathrm{max}}$ & $Z$\footnotemark[$\dagger$] 
    & EW$_{\mathrm{Fe\emissiontype{I}}}$ & $\chi _{\mathrm{red}}^{2}$ (dof) \\
    & (cm$^{-2}$) & (cm$^{-2}$) & & & & (keV) & ($Z_{\odot}$) & (eV) & \\
  \hline
  Averaged
    & $0.04$\footnotemark[$\ddagger$]
    & $\phantom{0}8.9_{-0.6}^{+0.5}$ & $0.75_{-0.02}^{+.002}$ 
    & $0.484$\footnotemark[$\ddagger$]
    & $1.00_{-0.07}^{+0.07}$ & $21.0_{-1.2}^{+1.3}$ & $2.10_{-0.10}^{+0.11}$ 
    & $65_{-\phantom{0}9}^{+\phantom{0}9}$ & $1.32$ ($860$) \\
  Phase~1
    & $0.04$\footnotemark[$\ddagger$]
    & $\phantom{0}8.8_{-0.8}^{+0.8}$ & $0.80_{-0.02}^{+0.02}$
    & $0.484$\footnotemark[$\ddagger$]
    & $0.97_{-0.12}^{+0.11}$ & $22.3_{-1.7}^{+1.9}$ & $2.04_{-0.15}^{+0.16}$
    & $63_{-13}^{+14}$ & $1.28$ ($617$) \\
  Phase~2
    & $0.04$\footnotemark[$\ddagger$]
    & $\phantom{0}7.1_{-1.2}^{+1.1}$ & $0.61_{-0.04}^{+0.03}$
    & $0.484$\footnotemark[$\ddagger$]
    & $1.15_{-0.12}^{+0.12}$ & $18.7_{-1.2}^{+1.8}$ & $1.99_{-0.16}^{+0.18}$
    & $52_{-18}^{+16}$ & $1.06$ ($486$) \\
  Phase~3
    & $0.04$\footnotemark[$\ddagger$]
    & $11.2_{-1.0}^{+1.0}$ & $0.83_{-0.03}^{+0.03}$
    & $0.484$\footnotemark[$\ddagger$]
    & $0.84_{-0.13}^{+0.13}$ & $23.5_{-2.2}^{+3.4}$ & $2.43_{-0.23}^{+0.25}$
    & $91_{-19}^{+21}$ & $1.25$ ($331$) \\
  \hline
  \multicolumn{9}{@{}l@{}}{\hbox to 0pt{\parbox{178mm}{\footnotesize
  \par \noindent
  \footnotemark[$*$]
    The parameters are
    the hydrogen column density of the interstellar medium ($N_{\mathrm{H}}^{\mathrm{ISM}}$)
    and that of the additional absorber ($N_{\mathrm{H}}^{\mathrm{CSM}}$),
    the covering fraction ($C^{\mathrm{CSM}}$), 
    the solid angle of the cold reflector viewed from the plasma ($\Omega /2\pi$),
    the power-law index of the emission measure ($\alpha$), 
    the maximum temperature of the plasma ($kT_{\mathrm{max}}$),
    the abundance of the plasma ($Z$),
    the equivalent width of 6.4~keV line (EW$_{\mathrm{Fe\emissiontype{I}}}$),
    the reduced $\chi$ squared ($\chi^{2}_{\mathrm{red}}$),
    and the degree of freedom (dof).
    The errors indicate the 90\% statistical uncertainty. \\ 
  \footnotemark[$\dagger$]
    We note that the abundance values are mainly determined from the iron lines,
    which are the most prominent lines, and an abundance of $Z_{\mathrm{Fe}}=2.0$ in
    \citet{Wilms2000} ([Fe/H]~$=2.69\times 10^{-5}$) is equivalent to that of $\sim$$1.1$
    in \citet{Anders1989} ([Fe/H]~$=4.68\times 10^{-5}$). \\ 
  \footnotemark[$\ddagger$]
    The values are fixed. 
  }\hss}}
\end{tabular}
\end{center}
\end{table*}
\begin{table*}[t]
\begin{center}
  \caption{Best-fit parameters for the ionized partial covering absorption model.\footnotemark[$*$]}
  \label{t3}
\begin{small}
\begin{tabular}{lcccccccccc}
  \hline
  Phase
    & $N_{\mathrm{H}}^{\mathrm{ISM}}/10^{21}$
    & $N_{\mathrm{H}}^{\mathrm{CSM}}/10^{21}$ & $C^{\mathrm{CSM}}$ & $\xi ^{\mathrm{CSM}}$
    & $\Omega /2\pi$
    & $\alpha$ & $kT_{\mathrm{max}}$ & $Z$\footnotemark[$\dagger$] 
    & EW$_{\mathrm{Fe\emissiontype{I}}}$ & $\chi _{\mathrm{red}}^{2}$ (dof) \\
    & (cm$^{-2}$) & (cm$^{-2}$) & & (erg~s$^{-1}$~cm) & & & (keV) & ($Z_{\odot}$) & (eV) & \\
  \hline
  Averaged
    & $0.04$\footnotemark[$\ddagger$]
    & $5.3_{-0.7}^{+0.5}$ & $0.82_{-0.02}^{+0.02}$ & $-0.62_{-0.15}^{+0.08}$
    & $0.484$\footnotemark[$\ddagger$]
    & $1.02_{-0.06}^{+0.06}$ & $18.6_{-0.4}^{+0.7}$ & $1.93_{-0.09}^{+0.10}$
    & $62_{-\phantom{0}9}^{+\phantom{0}8}$ & $1.31$ ($859$) \\
  Phase~1
    & $0.04$\footnotemark[$\ddagger$]
    & $5.5_{-1.4}^{+0.4}$ & $0.88_{-0.02}^{+0.02}$ & $-0.55_{-0.31}^{+0.02}$
    & $0.484$\footnotemark[$\ddagger$]
    & $0.98_{-0.12}^{+0.08}$ & $20.0_{-1.5}^{+1.8}$ & $1.90_{-0.16}^{+0.14}$
    & $60_{-15}^{+13}$ & $1.29$ ($616$) \\
  Phase~2 
    & $0.04$\footnotemark[$\ddagger$]
    & $4.3_{-2.3}^{+0.6}$ & $0.69_{-0.04}^{+0.04}$ & $-0.59_{-0.82}^{+0.11}$
    & $0.484$\footnotemark[$\ddagger$]
    & $1.10_{-0.10}^{+0.11}$ & $18.6_{-1.4}^{+1.8}$ & $1.93_{-0.17}^{+0.18}$
    & $50_{-16}^{+16}$ & $1.06$ ($485$) \\
  Phase~3
    & $0.04$\footnotemark[$\ddagger$]
    & $6.0_{-1.0}^{+1.3}$ & $0.89_{-0.03}^{+0.03}$ & $-0.71_{-0.22}^{+0.18}$
    & $0.484$\footnotemark[$\ddagger$]
    & $0.91_{-0.12}^{+0.14}$ & $20.2_{-2.2}^{+2.3}$ & $2.15_{-0.21}^{+0.23}$
    & $85_{-21}^{+21}$ & $1.22$ ($330$) \\
  \hline
  \multicolumn{11}{@{}l@{}}{\hbox to 0pt{\parbox{180mm}{\footnotesize
  \par \noindent
  \footnotemark[$*$], \footnotemark[$\dagger$], \footnotemark[$\ddagger$] 
    The notations and explanations follow table~\ref{t2} expect for
    the ionization parameter ($\xi ^{\mathrm{CSM}}$). 
  }\hss}} 
\end{tabular}
\end{small}
\end{center}
\end{table*}
%

\subsection{Spectral Analysis (2): Time-Sliced Spectra} \label{s4-4}
\begin{figure}[t]
\begin{center}
  \FigureFile(80mm,80mm){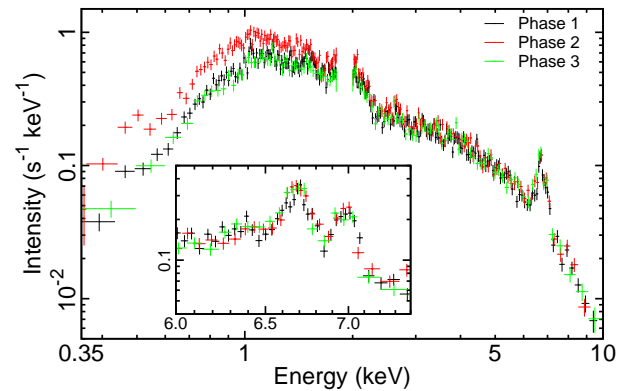}
\end{center}
  \caption{Background-subtracted BI spectra in the three phases (figure~\ref{f2})
           shown with different colors.
           The inset is an enlarged view for the Fe~K$\alpha$ complex lines using
           FI spectra.}
  \label{f8}
\end{figure}

We now proceed to time-sliced spectroscopy.
Figure~\ref{f8} shows the BI spectra in the three phases defined in figure~\ref{f2}. 
The elevated count rate in phase~2 in the soft band (0.2--2.0~keV) is confirmed in
the time-sliced spectra.
We applied the model constructed for the time-averaged spectra to the spectra of all
slices.
The best-fit parameters are shown in tables~\ref{t2} and \ref{t3}. 

\begin{figure*}[t]
\begin{center}
  \FigureFile(80mm,80mm){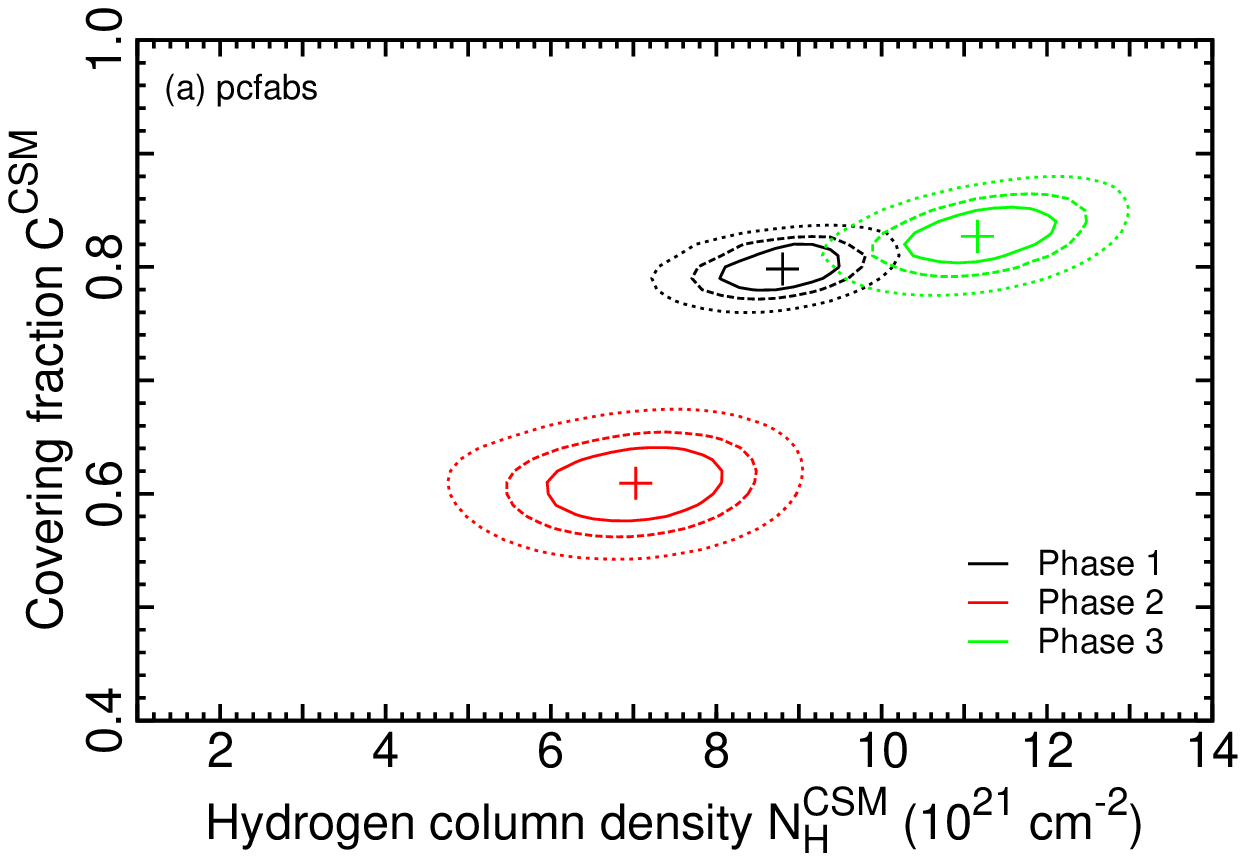}
  \FigureFile(80mm,80mm){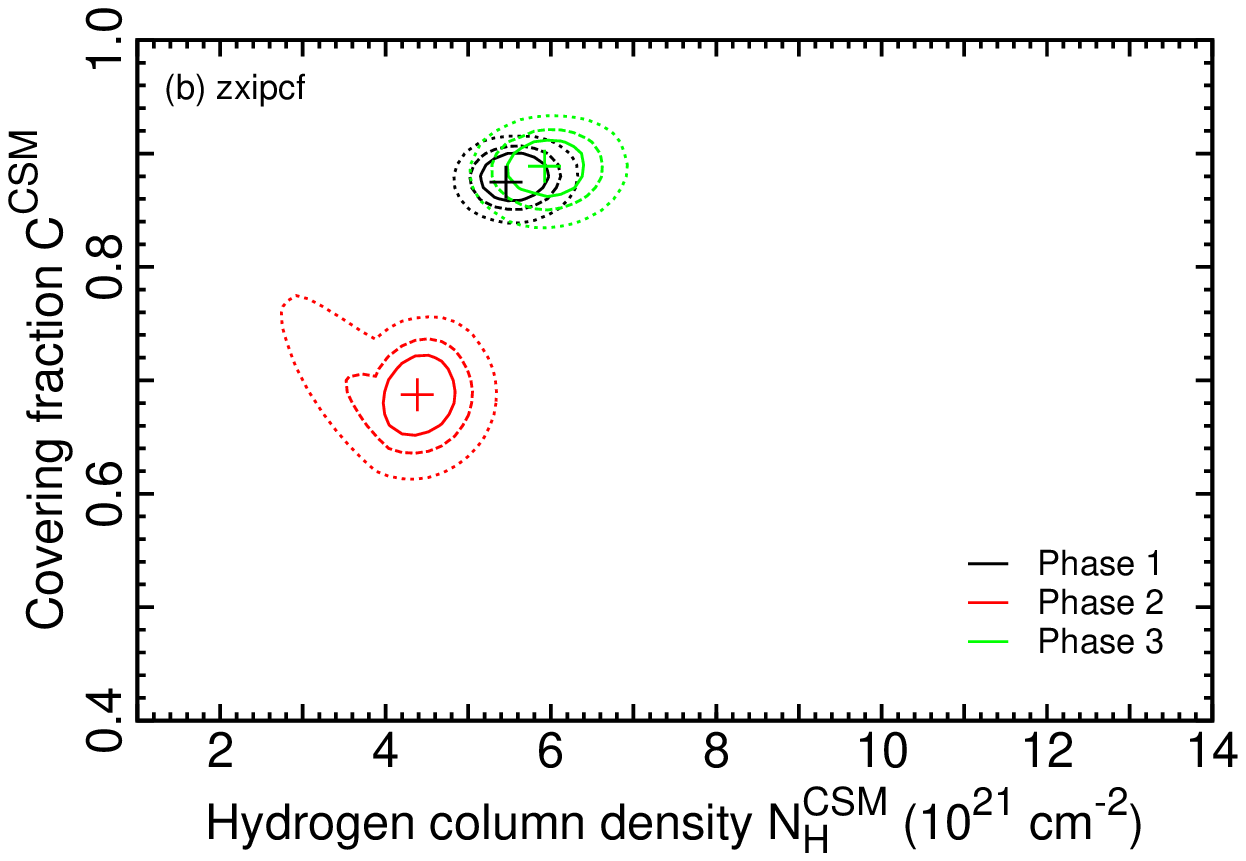}
  \FigureFile(80mm,80mm){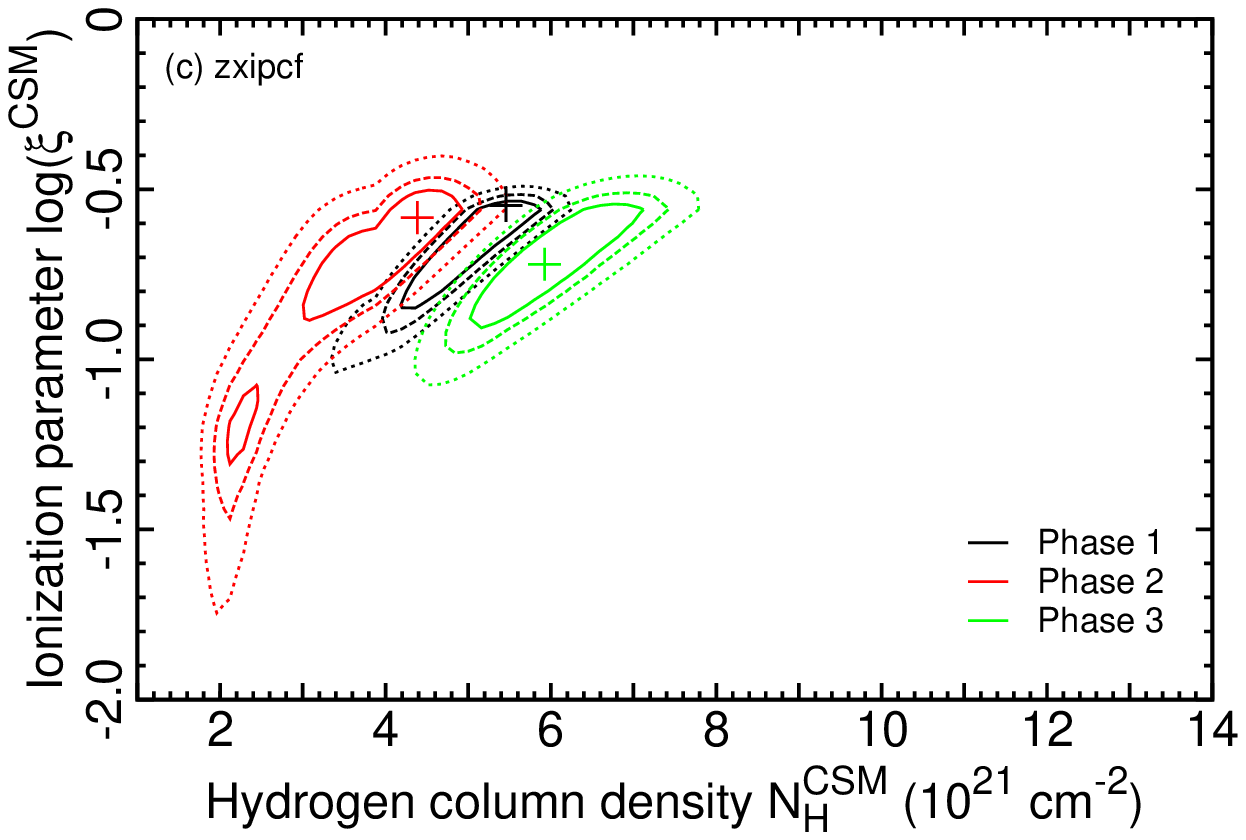}
  \FigureFile(80mm,80mm){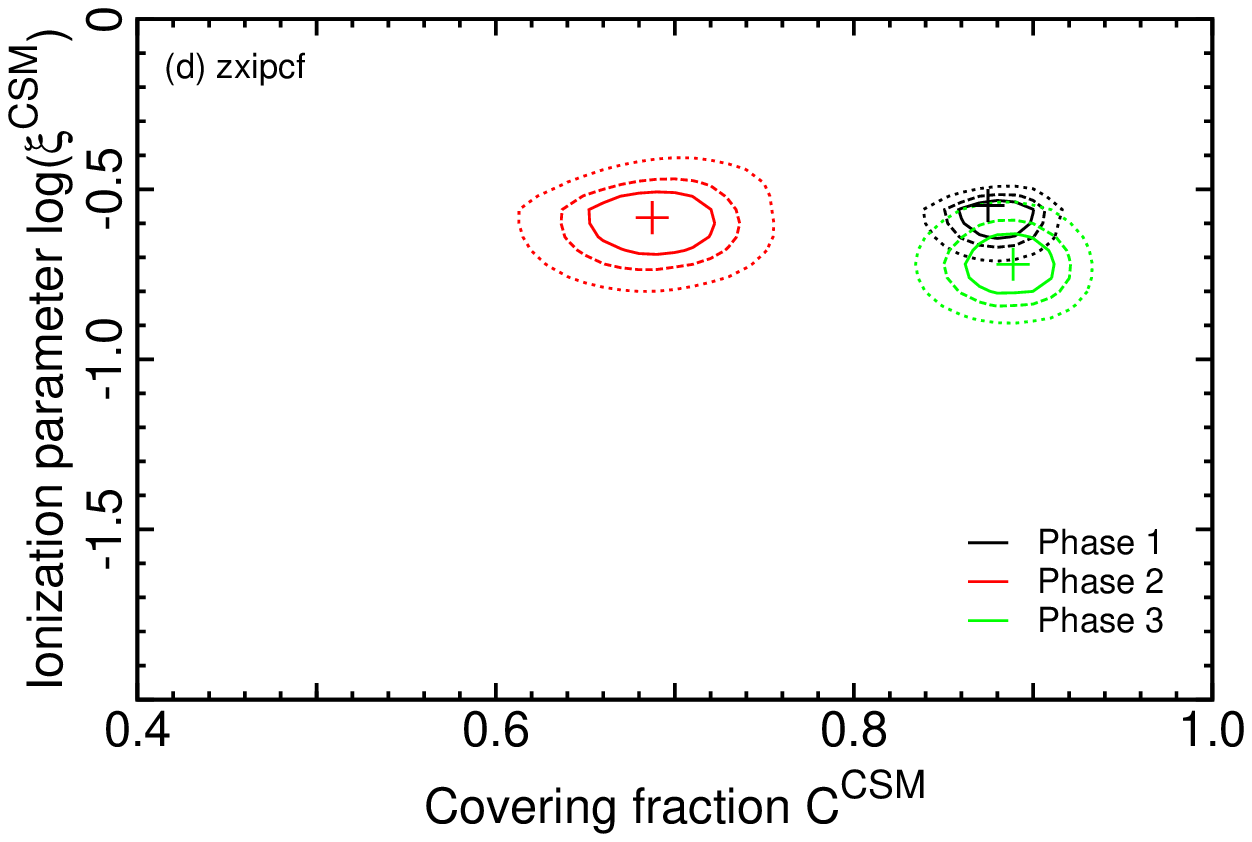}
\end{center}
  \caption{Contour plot for the best-fit parameters in the circumstellar absorption model.
           (a)~Partial absorption by neutral matter with the hydrogen column density
           ($N_{\mathrm{H}}^{\mathrm{CSM}}$) versus the covering fraction ($C^{\mathrm{CSM}}$).
           (b)~Partial absorption by ionized matter with $N_{\mathrm{H}}^{\mathrm{CSM}}$
           versus $C^{\mathrm{CSM}}$,
           (c)~with $N_{\mathrm{H}}^{\mathrm{CSM}}$ versus the ionization parameter
           ($\xi^{\mathrm{CSM}}$), and
           (d)~with $C^{\mathrm{CSM}}$ versus $\xi^{\mathrm{CSM}}$. 
           Each color represents parameters in each phase.
           The solid, dashed, and dotted curves indicate 1, 2, and 3$\sigma$ statistical
           confidence limits, respectively.
           The crosses indicate the best-fit values.}
  \label{f9}
\end{figure*}

The result of the fitting indicates that the time variation in phase~2 is attributable 
only to a change in the circumstellar absorption; other parameters do not change among 
the slices.
Figure~\ref{f9} shows the contour plots of the best-fit parameters in the circumstellar
absorption models.
A significant decrease of the covering fraction ($C^{\mathrm{CSM}}$) is seen in phase~2
in both models.
A hint of the change in the hydrogen column density ($N_{\mathrm{H}}^{\mathrm{CSM}}$) is
also seen in both models.
The ionization parameter ($\xi^{\mathrm{CSM}}$) in the ionized absorber model remains
constant among the three phases.
The spectral change in phase~2 is likely caused mainly by the change in the covering
fraction.

%

\section{Discussion} \label{s5}

\subsection{Suzaku Confirmation and Expansion of ASCA Results} \label{s5-1}
Z~Cam was observed twice with ASCA in a transition phase and an outburst phase
\citep{Baskill2001}.
Our Suzaku observation was the second to be made in the X-ray during the transition
phase of the object.
Our result confirm and expand the results presented in the preceding work, which we
briefly summarize here.
\citet{Baskill2001} clearly showed the presence of the circumstellar absorption, which
is larger than the interstellar absorption by two orders in the both the transition
and outburst phases.
The presence of such a large circumstellar absorption was confirmed by our observation
(tables~\ref{t2} and \ref{t3}).
\citet{Baskill2001} argued that the circumstellar absorption was ionized, but our data
were explained equally well by partially covering neutral material.

Upon the confirmation of the previous ASCA findings, we further presented some new
results, making use of the improved low-energy response and a wider band coverage of
Suzaku.
First, we found that the circumstellar absorption exhibits a time variation
(figures~\ref{f8} and \ref{f9}).
The variation is characterized by a decreased spectral hardness during a $\sim$20~ks
interval during our observation, which is mostly attributable to the change in the
covering fraction.
The variation is not associated with the orbital period of the system.
We also detected reprocessed emission of the Fe fluorescence and the Compton scattering
and constrained the plasma temperature and reflection geometry.

\subsection{Timing of the Suzaku Observation} \label{s5-2}
Our X-ray observation was made at the beginning of an optical outburst
(figure~\ref{f1}). The rapid rise and slow decay of the optical light curve suggest that
this outburst was an outside-in outburst (e.g., \cite{Lasota2001}), in which the heat
wave propagates from the outer part of the accretion disk to the inner part. However,
the X-ray characteristics are those typical of the quiescent phase in the following
three points.

First, in the Suzaku observation, the observed emission was quite hard with
a 2--10~keV luminosity at 163~pc (table~\ref{t1}) of
$8.5_{-0.1}^{+0.1}\times 10^{31}$~erg~s$^{-1}$.
The luminosity is quite similar to the EXOSAT measurement made at quiescence
\citep{Mukai1993} and is larger by 32 times than the ASCA measurement
made at an optical and presumably X-ray outburst phase \citep{Baskill2001}.
Here, we used the \texttt{pimms} tool\footnote{%
  See http://heasarc.gsfc.nasa.gov/docs/software/tools/pimms.html for details.
} for the flux conversion.
In the dichotomy of enhanced and suppressed hard X-ray emission in the quiescent
and outburst phases, respectively, the observed hardness and luminosity indicate
that the Suzaku observation was made during the X-ray quiescent phase.

Second, with the observed X-ray data, we can estimate the accretion rate from
the inner part of the disk to the WD surface ($\dot{M}_{\mathrm{BL}}$).
Assuming that an in-falling particle releases an energy of $(5/2)\,kT_{\mathrm{max}}$,
in which $kT_{\mathrm{max}}$ is the maximum plasma temperature of the boundary layer,
the bolometric luminosity of the boundary layer is described as
$L_{\mathrm{BL}}=(5\dot{M}_{\mathrm{BL}}\,kT_{\mathrm{max}})/(2\mu \,m_{\mathrm{H}})$, 
where $\mu$ is the mean molecular weight (typically $\sim$0.6) and $m_{\mathrm{H}}$
is the mass of hydrogen \citep{Fabian1994,Pandel2003}.
By substituting $L_{\mathrm{BL}}$ and $kT_{\mathrm{max}}$ with the 
observed values, we found that
$\dot{M}_{\mathrm{BL}}\sim 3\times 10^{-11}$~$\MO$~yr$^{-1}$,
which is comparable to the typical value during the quiescence of DNe 
\citep{Pringle1979,Pandel2003}, but is much smaller than the value required for
triggering outbursts ($\sim$$10^{-9}$~$\MO$~yr$^{-1}$: \cite{Osaki1996,Lasota2001}).

Third, we derived a constraint of the reflection geometry from the spectral
fitting using the hard X-ray emission above 20~keV and the Fe~K fluorescent line
\citep{Ishida2009}. 
Assuming that the WD surface contributes most as the reflector to reprocess
the X-rays from the plasma, the solid angle subtended by the reflector
($\Omega /2\pi$) converts to the scale height ($h$) of the X-ray plasma from
the WD surface (figure~\ref{f5}).
Using the observed value $\Omega /2\pi =0.484$ (tables~\ref{t2} and \ref{t3}),
$h\sim 0.17R_{\mathrm{WD}}=9.7\times 10^{7}$~cm (table~\ref{t1}), which is in line
with the boundary layer being the source of the X-ray emission as in other DNe
in the quiescent phase \citep{Mukai1997,Ishida2009}.

We thus conclude that our observation was made between the start of the state
change in the outer part of the disk and the arrival of the heat wave to the
inner part of the disk, in which optical outburst and X-ray quiescence co-exist.
Such a transition phase is expected to last for $\sim$1~day, which is the time
for the heating front to traverse the disk \citep{Lin1985}.
In fact, some X-ray observations were made in such a transition phase of DNe (e.g.,
\cite{Wheatley2003}), including one of the two ASCA observations of Z Cam
\citep{Baskill2001}.
The entire Suzaku observation was within this transition phase as we see no signs of
spectral change except for the change in the circumstellar absorption (\S~\ref{s4-3-3}
and \S~\ref{s4-4}), but the last part of the ASCA observation was in the optical
outburst phase as the X-ray luminosity declined in the last part of the observation
with no change of spectral hardness. 
It suggests that the phase for an outburst of our Suzaku observation was preceded
that of the ASCA observation.

\subsection{Cause for the Circumstellar Absorption} \label{s5-3}
We speculate the cause for the circumstellar absorption and its time variation.
Two possible agents for the absorption are (i)~disk wind and (ii)~geometrically
flaring disk.
Both models can explain the observed extinction feature modeled by a partially-covering
neutral or ionized absorption.
In the wind model, the ionized partial absorption is expected to intervene in the line
of sight with the covering fraction possibly representing the degree of porousness of
the wind.
In the flaring disk model, a part of the swollen disk is expected to cause a partial
coverage either by neutral or ionized matter.
The two models cannot be distinguished from our spectral results alone.
However, we also found a time-variation of the circumstellar absorption, which is not
associated with the orbital period.
We would naturally expect the variation associated with the orbital period in the
flaring disk interpretation, thus we favor the disk wind interpretation for the cause
of the circumstellar absorption. 
Figure~\ref{f5} shows a schematic view, in which the disk wind is triggered by the
propagation of the heat wave and causes a partial covering in the line of sight X-rays
from the plasma localized in the boundary layer.

What is perplexing is that the time variation of the circumstellar absorption is not
a monotonic increase and eventual saturation of the absorption column, which would be
na\"{i}vely expected in the disk wind interpretation.
The variation that we saw might be due to the change in the degree of porousness of
the relatively stable wind, rather than the change in strength of the wind.
A longer coverage of an outburst phase as well as the coverage of other phases far from
the outburst and transition phases with the same data quality will help us to test our
ideas.

The mechanism to launch disk wind is not well understood yet.
If the absorption seen in our observation is caused by the wind, we argue that X-ray
radiation pressure does not play an important role at least in the X-ray quiescent phase.
First, the increase of the additional absorption in phase~2 does not accompany any changes
in the intrinsic X-ray emission properties.
Second, the X-ray radiation energy absorbed by the additional absorber can be derived as
a difference between the additional-absorption corrected and uncorrected luminosity in
the 0.35--10~keV band.
The value $\sim$$3\times 10^{31}$~erg~s$^{-1}$ is smaller than the wind mechanical energy
of $\sim$$8\times 10^{32}$~erg~s$^{-1}$ by more than an order with a mass loss rate of
$\ge$$2.4\times 10^{-9}$~\MO~yr$^{-1}$ \citep{Robinson1973} and a mean wind velocity of
1000~km~s$^{-1}$ (e.g., \cite{Warner1995}).

\section{Summary} \label{s6}
We conducted a Suzaku X-ray observation of the dwarf nova Z~Cam at the onset of an
optical outburst by chance.
The X-ray spectral characteristics, however, suggest that the source was in the X-ray
quiescent phase.
This implies that our observation was made at the time when the heat wave had not
reached to the inner part of the accretion disk in the development of the outburst,
during which optical outburst and X-ray quiescent phases co-exist.

The X-ray spectrum shows clear evidence of an extra absorption upon the interstellar
absorption, as was presented in the previous work.
We found that
(i)~the extra absorption was modeled successfully by partial coverage either by neutral
or ionized matter,
(ii)~the absorption shows a time variation with a time scale of $\sim$20~ks,
(iii)~the variation is mostly attributable to the change in the covering fraction, and
(iv)~the variation is not associated with the orbital period of the system.
From these finding, we argued that the circumstellar absorption can be either by disk
wind or geometrically flaring disk that intervene in the line of sight of the X-rays
located in the boundary layer between the WD surface and the disk.

\bigskip

The authors acknowledge the referee, Knox~S.~Long, for improving the paper. 
The authors appreciate Dai~Takei for his help in Suzaku data analysis. 
K.\,S. is financially supported by the Japan Society for the Promotion of Science.
We acknowledge the variable star observations from the AAVSO International Database
contributed by observers worldwide. 
This research made use of data obtained from Data ARchives and Transmission
System (DARTS), provided by Center for Science-satellite Operation and Data
Archives (C-SODA) at ISAS/JAXA.


\end{document}